\documentclass[twocolumn,twocolappendix]{aastex701}

\usepackage{physics}
\usepackage{CJK}
\usepackage{comment}
\usepackage{enumitem}
\usepackage{xcolor}

\newcommand{\E}{\mathcal{E}}
\newcommand{\F}{\mathbf{F}_{r}}
\newcommand{\G}{G^{0}}

\newcommand{\eg}{e_{\rm{g}}}

\newcommand{\M}{M_{\star}}
\newcommand{\aorb}{a_{\rm{orb}}}

\begin{document}
\begin{CJK*}{UTF8}{gbsn}

\title{Unravelling the Nature of AT\,2025abao: From Precursor to Final Fate}

\author[orcid=0000-0001-7420-9606]{Zhuo Chen (陈卓)}
\affiliation{Chen-Ning Yang Institute for Advanced Study, Tsinghua University, Beijing 100084, P.R. China}
\email[show]{chenzhuo\_astro@tsinghua.edu.cn}

\author[orcid=0000-0002-0029-8682]{Yingzhen Cui (崔英朕)} 
\affiliation{National Astronomical Observatories, Chinese Academy of Sciences, Beijing 100101, P.R. China}
\email[show]{\\cuiyz@bao.ac.cn}

\author[orcid=0009-0008-1033-5476]{Tianshu Wu (吴天枢)}
\affiliation{Department of Astronomy, Tsinghua University, Beijing 100084, P.R. China}
\email{wuts25@mails.tsinghua.edu.cn}

\author[orcid=0000-0002-7714-493X]{Yongzhi Cai (蔡永志)}
\affiliation{Yunnan Observatories, Chinese Academy of Sciences (CAS), Kunming 650216, P.R. China}
\affiliation{INAF -- Osservatorio Astronomico di Padova, Vicolo dell'Osservatorio 5, 35122 Padova, Italy}
\email[show]{yongzhi.cai@inaf.it}

\author[orcid=0000-0002-6398-0195]{Hongwei Ge (葛宏伟)}
\affiliation{Yunnan Observatories, Chinese Academy of Sciences (CAS), Kunming 650216, P.R. China}
\affiliation{International Centre of Supernovae, Yunnan Key Laboratory of Supernova Research, Yunnan Observatories, Chinese Academy of Sciences, Kunming, 650216, P.R. China}
\affiliation{University of Chinese Academy of Sciences, Beijing 100049, P.R. China}
\email{gehw@ynao.ac.cn}

\author[orcid=0000-0003-4254-2724]{Andrea Reguitti}
\affiliation{INAF -- Osservatorio Astronomico di Padova, Vicolo dell'Osservatorio 5, 35122 Padova, Italy}
\email{andrea.reguitti@inaf.it}

\author[orcid=0000-0003-4721-6477]{Minxuan Cai}
\affiliation{Department of Astronomy, University of Science and Technology of China, Hefei 230026, P.R. China}
\email{caimx@mail.ustc.edu.cn}

\author[orcid=0000-0003-4200-4432]{Lulu Fan (范璐璐)}
\affiliation{Department of Astronomy, University of Science and Technology of China, Hefei 230026, P.R. China}
\affiliation{School of Astronomy and Space Science, University of Science and Technology of China, Hefei 230026, P.R. China}
\affiliation{College of Physics, Guizhou University, Guiyang 550025, P.R. China}
\affiliation{Deep Space Exploration Laboratory, Hefei 230088, P.R. China}
\email{llfan@ustc.edu.cn}

\author[orcid=0000-0002-9092-0593]{Ji-An Jiang (姜继安)}
\affiliation{Department of Astronomy, University of Science and Technology of China, Hefei 230026, P.R. China}
\affiliation{National Astronomical Observatory of Japan, National Institutes of Natural Sciences, Tokyo 181-8588, Japan}
\email{jian.jiang@ustc.edu.cn}

\author[orcid=0000-0002-1517-6792]{Tinggui Wang (王挺贵)}
\affiliation{Department of Astronomy, University of Science and Technology of China, Hefei 230026, P.R. China}
\email{twang@ustc.edu.cn}

\author[orcid=0009-0009-5238-2059]{Zelin Xu (徐则林)}
\affiliation{Department of Astronomy, University of Science and Technology of China, Hefei 230026, P.R. China}
\email{sa21022025@mail.ustc.edu.cn}

\author[orcid=0000-0003-3965-6931]{Jie Lin (林杰)}
\affiliation{Department of Astronomy, University of Science and Technology of China, Hefei 230026, P.R. China}
\email{linjie2019@ustc.edu.cn}

\begin{abstract}
Luminous red novae (LRNe) are potential signatures of common envelope evolution (CEE), a critical yet poorly understood phase in compact binary formation. AT\,2025abao, an LRN in the Andromeda galaxy, originated from an asymptotic giant branch (AGB) progenitor and exhibited a prolonged pre-outburst transient. Lacking direct constraints on the companion, however, the system's evolutionary fate remains unclear. Here, we unravel the binary nature of AT\,2025abao by modeling its complete evolution, from precursor to final outcome. We show that the eight-year infrared precursor arises from a $\approx1.4^{+0.2}_{-0.3}\,M_{\odot}$ main-sequence companion inspiralling through the wind of a $\approx 6.9\,M_{\odot}$ AGB star. As CEE commences, the companion plunges into the envelope and launches a shock. Radiation hydrodynamic modeling of this outburst constrains the total ejecta mass to $\approx 0.92\,M_{\odot}$ and energy to $\approx 1.25\times 10^{47}\,$erg, revealing that only $\approx 0.57\,M_{\odot}$ becomes unbound. Subsequent radiative transfer simulations successfully reproduce the evolving spectral energy distribution. Crucially, our models indicate the main-sequence companion and AGB core survive in an eccentric orbit without merging at least 400 days after CEE onset. AT\,2025abao may therefore be the first observed CEE event caught forming a close compact binary, bridging a long-standing observational gap in compact binary formation.
\end{abstract}


\section{Introduction}

Common envelope evolution (CEE) is one of the most mysterious and important stages in binary evolution \citep{paczynski1976,ivanova2013,ropke2023}. Supernovae \citep[SNe; ][]{liu2023}, kilonovae \citep{metzger2010}, X-ray binaries \citep{kalogera1998}, pulsars \citep{yang2025}, hot subdwarfs \citep{han2002,lin2023}, and planetary nebulae \citep{jones2017} may all have undergone one or more CEE phases. However, CEE remains poorly understood owing to the scarcity of direct observations and the complexity of theoretical modeling. A widely accepted picture is that luminous red novae (LRNe), classified as gap transients \citep{cai2022a}, could be an observational signature of CEE \citep{kaminski2026}. The link between CEE and LRNe was solidified by the event V1309 Sco, whose light curve and decaying orbital period were captured by OGLE. Follow-up analysis revealed that its progenitor was likely a $\approx1.5\,M_{\odot}$ main-sequence (MS) star with a $\approx0.1\,M_{\odot}$ companion \citep{tylenda2011}.

However, as a main-sequence merger, V1309 Sco cannot answer the question of how close compact binaries form through CEE, and it remains the only LRN with unambiguous evidence of a binary origin, despite the discovery of many more LRNe with advanced telescopes \citep{blagorodnova2017,cai2019,pastorello2019,blagorodnova2021,pastorello2021a,pastorello2021b,cai2022b,reguitti2026a}. The key challenge in characterizing the nature of LRNe stems from a lack of well-sampled pre-outburst data, high-quality follow-up observations during the outburst, and long-term monitoring of the merger remnants. Among all LRNe, the recent event AT\,2025abao in the Andromeda galaxy (M31) stands out for its quiescent asymptotic giant branch (AGB) progenitor and its precursor observations in the infrared and optical domains. In this work, we unravel the binary nature of AT\,2025abao by modeling its complete evolution, from precursor to final fate.

\section{AT\,2025abao observations}

The quiescent progenitor of AT\,2025abao was identified as an AGB star with a mass $M_{1}\approx7\pm2\,M_{\odot}$, $\log_{10}(L_{\rm AGB}/L_{\odot})\approx4.2\pm0.1$, and $R\approx350\pm50\,R_{\odot}$ \citep{karambelkar2025}. The precursor of AT\,2025abao, named WNTR23bzdiq/WTP19aalzlk, gradually brightened from $L\approx16000\,L_{\odot}$ in 2018 to $L\approx50000\,L_{\odot}$ in 2025 \citep{karambelkar2025}, as identified by the Wide-field Infrared Transient Explorer and NEOWISE surveys.

On 2025 November 2, AT\,2025abao was firmly classified as an LRN by \citet{taguchi2025}, based on its spectral similarities with another LRN, AT\,2019zhd \citep{pastorello2021a}, also in M31.
The onset of the LRN outburst was characterized by a rapid $r$-band brightening of approximately five magnitudes within one week following the discovery of AT\,2025abao. The outburst lasted for about 90 days, exhibiting a plateau in the light curve with a nearly constant luminosity (i.e., $3$--$4\times10^{39}$\,erg\,s$^{-1}$), before gradually declining to $\approx10^{39}$\,erg\,s$^{-1}$ by 2026 March. The follow-up campaign for AT\,2025abao was suspended at this stage due to solar conjunction. Details regarding the photometric observations and data reduction are provided in Appendix \ref{subsec:photometry}. In addition to the photometric evolution, spectroscopic observations during the outburst revealed an H$\alpha$ full width at half maximum (FWHM) velocity of $\approx900$\,km\,s$^{-1}$ \citep{reguitti2026b}. This indicates the presence of high-velocity ejecta with a possible magnetohydrodynamic origin \citep{ondratschek2022,vetter2024,vetter2025}.

\section{Precursor in the wind}

Although direct observational evidence for a companion star is lacking, the eight-year gradual brightening fits a slow-inspiral scenario within a high-density AGB wind. In this model, the companion slowly loses angular momentum to wind drag \citep{ostriker1999}, driving orbital decay and releasing gravitational energy. The drag force converts this gravitational energy into internal energy, which efficiently dissipates via radiation transport due to the wind's low optical depth. To first order, the excess luminosity $\dot{E}_{\rm heat}$ from the inspiral is proportional to the gravitational drag force $F_{\rm d}$. Additionally, because the wind removes angular momentum from the binary via this drag, assuming the specific angular momentum carried by the wind is comparable to that at the L2 point establishes an upper limit on the drag force, linking it directly to the wind mass-loss rate \citep{pejcha2017}. We detail this analytic drag-in-the-wind model in Appendix \ref{app:wind}.

\begin{figure*}[!htb]
    \centering
    \includegraphics[width=\textwidth]{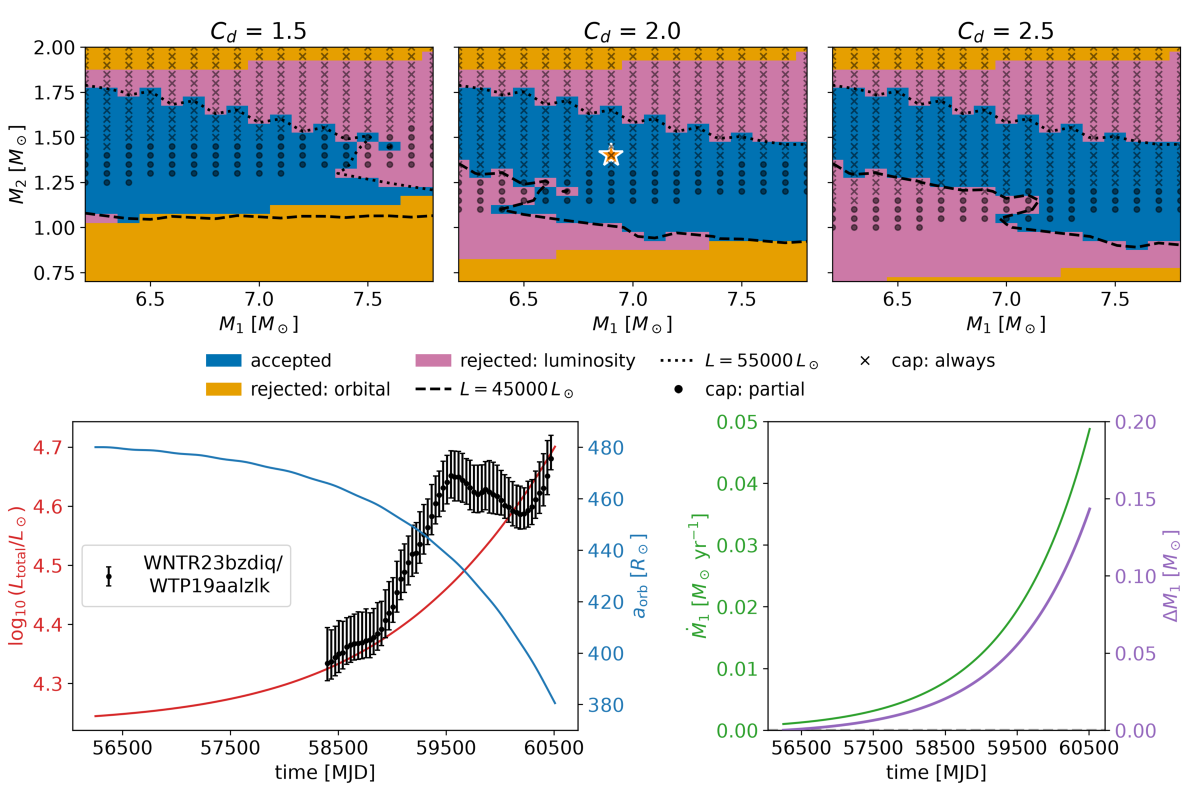}
    \caption{First row: compatible binary mass combinations predicted by the analytic drag-in-the-wind model (Appendix \ref{app:wind}), with the representative binary model ($M_{1}=6.9\,M_{\odot}$, $M_{2}=1.4\,M_{\odot}$) studied throughout this work indicated by the orange star. Symbols $\times$ and $\bullet$ denote models whose $F_{\rm d}$ are limited by Equation \ref{eqn:fdlimit} entirely or partially during the evolution, respectively. Second row: the evolution of the orbit, total luminosity (defined in Appendix \ref{app:wind}), wind mass-loss rate, and cumulative mass loss of the representative simulation.}
    \label{fig:precursor}
\end{figure*}

Using this drag-in-the-wind model, an assumed initial binary separation $\aorb \approx 480\,R_{\odot}$, and an enhanced AGB wind (discussed in Appendix \ref{app:wind}), we constrain the companion mass based on the precursor's maximum luminosity and orbital evolution. We conduct a parameter survey to identify viable progenitor binary combinations.

In the top three panels of Figure \ref{fig:precursor}, blue regions indicate compatible companion masses across different drag coefficients $C_d$ and primary masses $M_1$. Pink regions denote rejected parameters where the predicted maximum luminosity deviates by over 10\% from observations. Yellow regions represent systems where the binary separation remains $a_{\rm orb} > 380\,R_{\odot}$ after the wind ejects $0.2\,M_{\odot}$, indicating the orbital decay is too slow for the binary to initiate CEE in time. Crosses ($\times$) mark models where the drag force is constrained by the angular momentum loss rate limit (Equation \ref{eqn:fdlimit}) throughout the evolution, while bullets ($\bullet$) indicate models where this limit applies only partially. The remaining binary combinations never reach this limit and are calculated using the original drag force formula from \citet{bronner2024}.

Overall, the predicted companion mass, $M_2 \approx 1.4^{+0.2}_{-0.3}\,M_{\odot}$, depends only weakly on $M_1$ and $C_d$. Because $\dot{E}_{\rm heat}$ is generally positively correlated with both $M_1$ and $M_2$, these masses are negatively correlated at a fixed total luminosity $L_{\rm total}$. A representative simulation ($C_d = 2$, $M_1 = 6.9\,M_{\odot}$, $M_2 = 1.4\,M_{\odot}$) is shown in the bottom panels of Figure \ref{fig:precursor}, comparing its total luminosity against the observationally inferred bolometric luminosity, alongside its orbital separation, wind mass-loss rate, and cumulative mass loss. The smooth total luminosity of this representative model inherently struggles to reproduce the observed luminosity bump around MJD 59500. This feature likely requires additional physical mechanisms, such as a giant eruption \citep{pastorello2019} or the engulfment of a planetary-mass third body \citep{de2023,lau2025b,lau2025c}. In this model, the total mass lost during the enhanced-wind phase is $\approx 0.15\,M_{\odot}$.

\section{CEE modeled with {\tt MESA}}\label{sec:mesa}

\begin{figure*}[!htb]
    \centering
    \includegraphics[width=\textwidth]{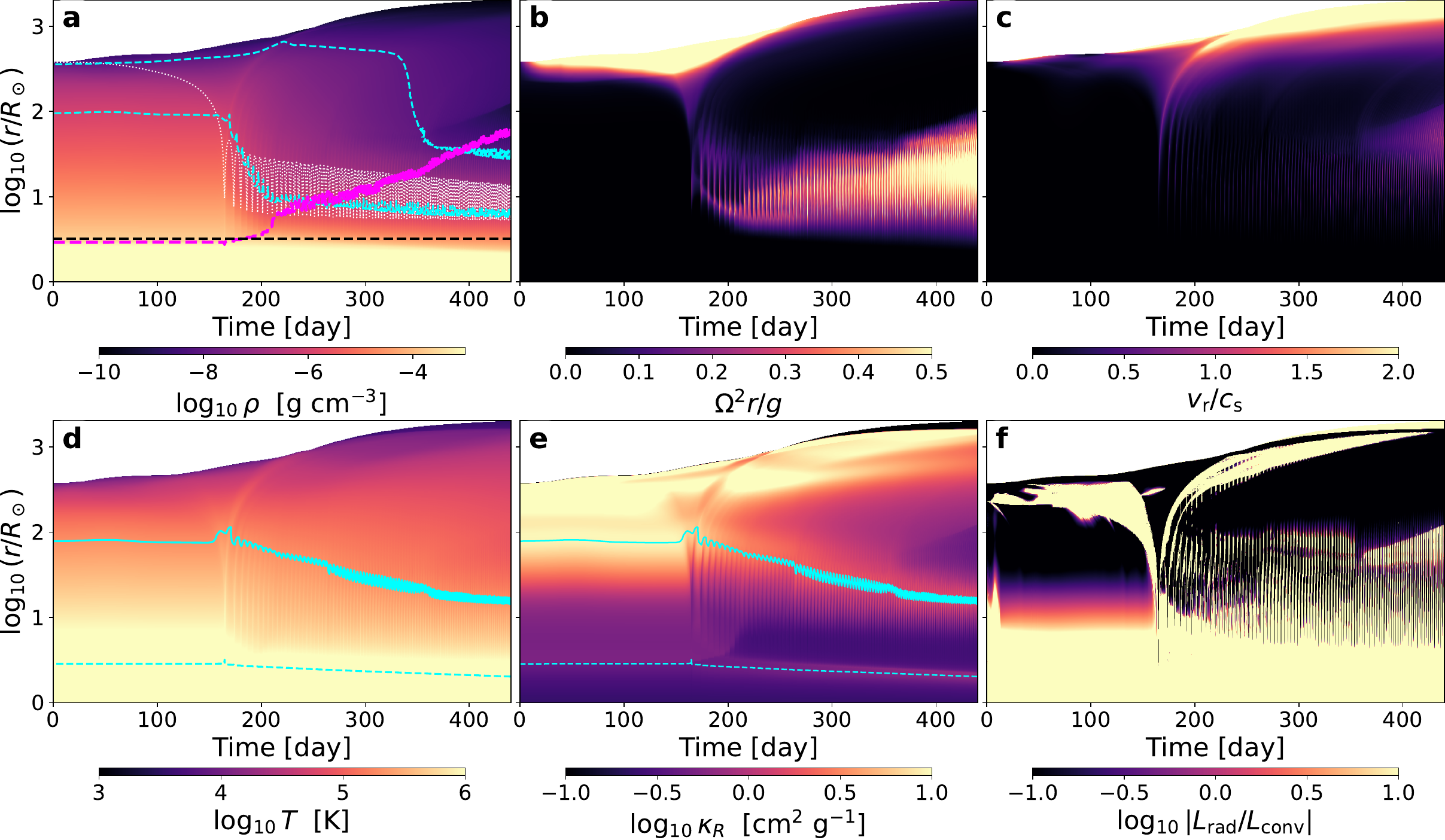}
    \caption{{\tt MESA} simulation of the representative binary model, using the angular-momentum-conserving framework (Appendix \ref{app:mesa}). Panels a--f show the evolution of the density, the ratio of centrifugal force to gravity, the ratio of radial velocity to local sound speed, the temperature, the Rosseland mean opacity, and the ratio of the local radiative to convective luminosity. In panel a, the cyan dashed lines mark $\rho=\{10^{-8},10^{-6}\}\rm{g}\, \rm{cm}^{-3}$; the white dotted and black dashed lines show the binary separation and the tidal disruption separation; the pink line shows the $90\%$ helium mass-fraction contour, marking the boundary of the helium envelope. The cyan solid lines in panels d and e mark $T=\{2\times10^{5},2\times10^{6}\}\,$K, the two opacity bumps arising from iron-group transitions.}
    \label{fig:cee1d}
\end{figure*}

As the binary separation decreases further, CEE begins, and the companion star rapidly plunges into the envelope due to gravitational drag. To model the complex physics of CEE, we develop a one-dimensional, energy- and angular-momentum-conserving model (Appendix \ref{app:mesa}) implemented in {\tt MESA} r10398 \citep{mesa2011,mesa2013,mesa2015,mesa2018,mesa2019,mesa2023}. We initialize the binary with companion masses selected from the compatible (blue) regions in Figure \ref{fig:precursor} (see Appendix \ref{app:mesainit}). While extensive parameter space exploration yields qualitatively similar results, we focus on the representative simulation with $M_{1}=6.9\,M_{\odot}$, $M_{2}=1.4\,M_{\odot}$, and $C_d=1$ (no wind focusing), shown in Figure \ref{fig:cee1d}. Initially, the companion is placed at $\aorb=380\,R_{\odot}$, coinciding with the end state of the drag-in-the-wind simulation. Because its orbital frequency exceeds the envelope's rotation frequency, it experiences a strong drag force. This drag transfers orbital angular momentum to the envelope and converts the companion's kinetic and gravitational energy into the envelope's internal energy, driving a rapid inspiral (Figure \ref{fig:cee1d}, white dotted line). This initial plunge ends after $\sim170$ days, with the companion gradually settling into an eccentric orbit from $\sim180$ days onward. Simultaneously, excess internal and recombination energy drives envelope expansion, consistent with previous studies \citep{fragos2019,bronner2024}. Significant centrifugal force reduces the envelope's effective gravity, further enhancing this expansion (Figure \ref{fig:cee1d}b). Iron-group bound-bound and bound-free transitions create an opacity valley between $2\times10^{5}\,$K and $2\times10^{6}\,$K (Figure \ref{fig:cee1d}d and \ref{fig:cee1d}e, cyan lines). Coupled with a sharp density drop between $10$--$100\,R_{\odot}$, radiative transfer becomes highly efficient in this region (Figure \ref{fig:cee1d}f). This radiative zone may inhibit convective angular momentum transport, helping to sustain the strongly rotating layer.

During the plunge-in phase (0--170 days), the companion travels supersonically within the envelope. Physically, the localized energy deposition could produce a radiation energy dominated shock \citep{macleod2017,gagnier2026} that can propagate to the envelope's surface. In Figure \ref{fig:cee1d}c, a supersonic shock is visible from $\sim180$\,days and travels to the surface of the envelope around 230 days. We propose that the observed outburst is the observational signature of this shock breaking out through the envelope's surface. However, {\tt MESA} simulations with a spherically symmetric approximation cannot resolve this highly dynamical and localized shock; the energy deposition range, controlled by $C_h$ (see Appendix \ref{app:ch}), may also be too wide compared to a realistic shock. In this simulation, the shock is dominated by a hydrodynamically dominated shock, with the maximum radiation-to-gas pressure ratio around 0.25. Future 2D/3D RHD models may better resolve the physical properties of the shock \citep{gagnier2026}.

As the companion dives deeper and the envelope expands, the enclosed mass within the impact radius drops. This weakens the gravitational drag, slowing the inspiral \citep{hirai2022} and transitioning the CEE into a self-regulated phase \citep{ivanova2016,gagnier2025}. The outcome of this phase depends on the binary properties and subsequent angular momentum transport. The main-sequence companion has a radius $R_{2}\approx1.38\,R_{\odot}$. The AGB core comprises a C/O core ($M_{\rm CO}\approx1\,M_{\odot}$, $R_{\rm CO}\approx0.3\,R_{\odot}$) surrounded by a helium envelope ($M_{\rm He}\approx0.8\,M_{\odot}$, $R_{\rm He}\approx2.93\,R_{\odot}$). The slowly inspiralling companion shrinks the Roche lobe of the AGB helium envelope, eventually stripping it near periastron (Figure \ref{fig:cee1d}a, pink 90\% helium mass-fraction contour). Because the helium envelope's mass, moment of inertia, and density are significantly lower than those of the companion, subsequent inspiral driven purely by gravitational drag remains inefficient. By the simulation's end ($400$ days), the companion has not reached its tidal disruption radius relative to the C/O core (black dashed line). While our 1D simulation considers only gravitational drag, two other angular momentum transfer mechanisms likely operate during this self-regulated phase. First, 3D magnetohydrodynamic simulations indicate that the magneto-rotational instability (MRI) \citep{balbus1991} can amplify the envelope's magnetic field and efficiently extract orbital angular momentum over secular timescales \citep{gagnier2024,vetter2025}. Second, the helium envelope and the companion's eccentric orbit may excite strong dynamical tides \citep{lai1997}.

We also performed simulations with $M_{2}\in\{1.1,1.2,1.3,1.5,1.6\}\,M_{\odot}$. These models exhibit similar evolutionary trajectories. As physically expected, a less massive companion results in a smaller final separation and a less expanded envelope.

\section{From shocked ejecta to LRN modeled with {\tt Guangqi}}

\begin{figure*}[!htb]
    \centering
    \includegraphics[width=\textwidth]{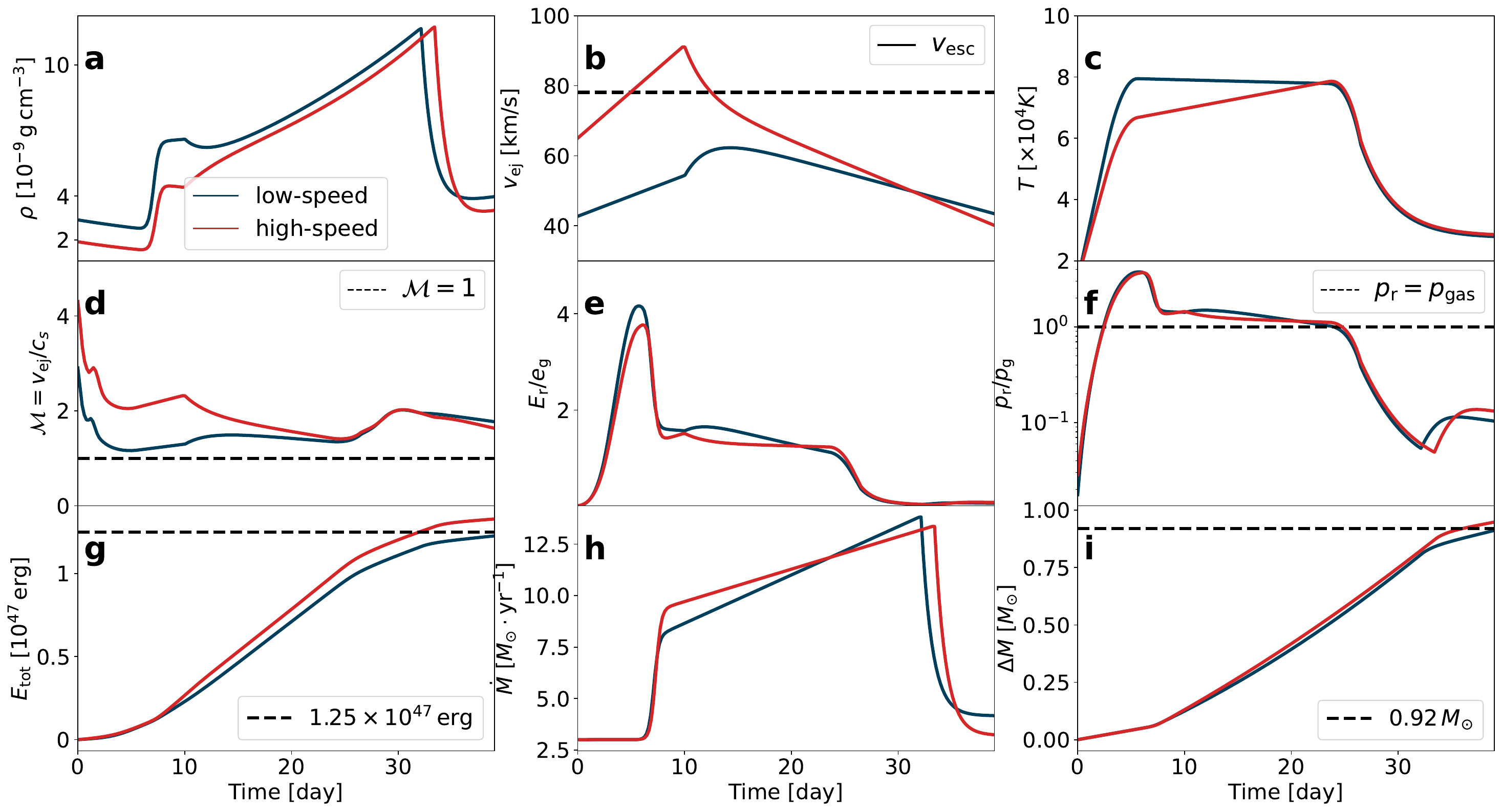}
    \caption{Two representative ejecta models whose bolometric luminosities fit the light curve of AT\,2025abao. Panels a--i show the time-dependent density, ejecta radial velocity, temperature, Mach number, $E_{r}$-to-$\eg$ ratio, $p_{r}$-to-$p_{g}$ ratio, cumulative gas and radiation energies, mass injection rate through the inner boundary, and cumulative injected mass. The two ejecta have very different initial speeds but similar cumulative masses and energies.}
    \label{fig:boundcond}
\end{figure*}

\begin{figure*}[!htb]
    \centering
    \includegraphics[width=\textwidth]{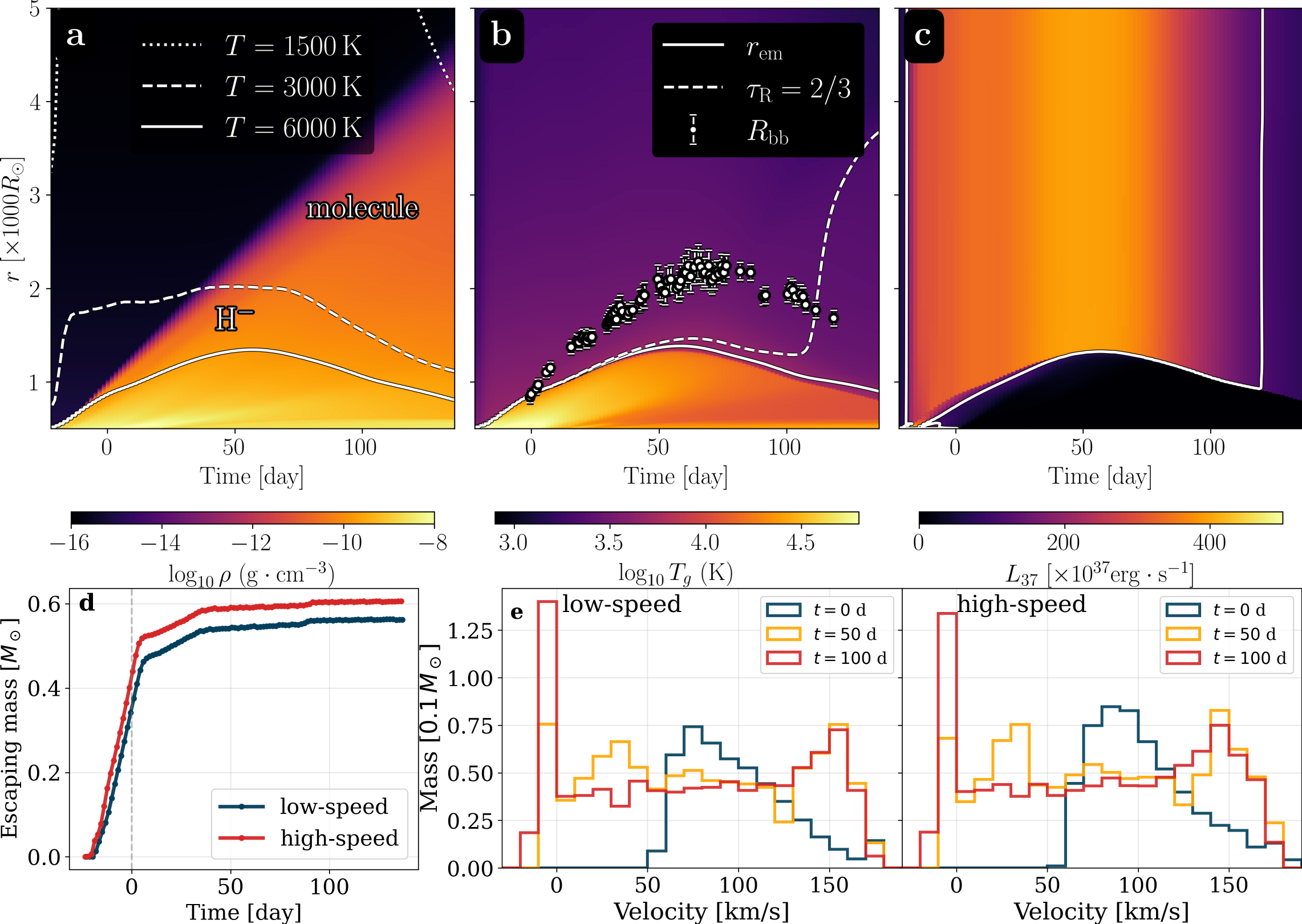}
    \caption{Evolution of the low-speed ejecta, shifted in time to fit the observed bolometric light curve. Panels a--c show the density, temperature, and radiative luminosity, respectively. The density panel is overplotted with $T=\{1500,3000,6000\}\,$K contours. The temperature panel is overplotted with the conventional photospheric surface $\tau_{R}=2/3$, the photon emission sphere $r_{\rm em}$ (defined in Appendix \ref{app:radmc3d}), and the blackbody radius fitted to the multi-wavelength observations. The white line in panel c shows $L_{\rm Edd}$, based on the electron-scattering opacity $\kappa_{R}=0.34\,\rm{cm}^{2}\, \rm{g}^{-1}$. Panel d: escaping mass of the RHD ejecta. Panel e: the ejecta velocity distributions of both low- and high-speed solutions at $t=\{0,50,100\}$~days.}
    \label{fig:evolution}
\end{figure*}

When the companion-induced shock reaches the envelope surface, it can eject material and produce a light-curve outburst resembling energetic transients like electron-capture \citep{karambelkar2023,wang2026} and failed supernovae \citep{andrea2025}, but preferentially in the orbital plane \citep{kirilov2025} and with lower energy. To model the ejecta that reproduces the AT\,2025abao light curve, we employ the improved radiation-hydrodynamic (RHD) code {\tt Guangqi}\footnote{The working code is available at \url{https://github.com/czlovemath123/guangqi1d} while the version used in this work is in Zenodo at \dataset[doi: 10.5281/zenodo.23175683]{https://doi.org/10.5281/zenodo.23175683}.} \citep{chen2026a,chen2026b}. We also refine the light-curve fitting using a Gaussian process (GP) and expected improvement (EI) framework (Appendix \ref{app:lcfit}). This 1D spherical RHD model solves the radiation transport equation via the flux-limited-diffusion approximation \citep{levermore1981,levermore1984} using a grey opacity table from \citet{malygin2014}. The hydrodynamic equations incorporate an analytic equation of state with mass fractions $X=0.74$ and $Y=0.26$ \citep{chen2019}. Our fitting functions assume: a fixed central mass $M=8\,M_{\odot}$; a fixed inner boundary for ejecta emergence at $R_{\rm in}=500\,R_{\odot}$; and a parameterized mass injection rate $\dot{M}$, temperature $T$, and velocity $v_{\rm ej}$. These variables rise to a peak and subsequently decline, mimicking the outburst light curve. The bolometric light curve is approximated by the radiation energy flux at the outer boundary $R_{\rm out}=10000\,R_{\odot}$.

The GP and EI method identifies a set of inner boundary conditions (``solutions'') that fit the bolometric light curve. Figure \ref{fig:boundcond} presents two representative ejecta solutions---a purely sub-escape profile and a higher-speed profile---that bracket a continuous spectrum of valid velocity profiles. The energy conversion between the kinetic energy and radiation energy contributes to this velocity spread, with the low-speed profile starting with higher radiation energy. In both simulations, radiation pressure $p_{\rm r}$ in the ejecta briefly exceeds gas pressure $p_{\rm g}$, placing the material in a radiation-pressure-driven regime \citep{chen2024,chen2026a}. The discrepancy in $p_{\rm r}/p_{\rm g}$ between these solutions and the {\tt MESA} simulations may arise from 1D limitations and the assumed energy deposition mechanism, as discussed in Section \ref{sec:mesa}. Though not unique, these solutions share key features. Each solution yields a total ejecta energy of $\approx 1.25\times10^{47}\,$erg and an injected mass of $\approx 0.92\,M_{\odot}$. A fraction of this mass escapes, falling well below the $\approx 5\,M_{\odot}$ total envelope mass. Therefore, the light-curve fitting implies the envelope remains partially bound after CEE. Encouragingly, the {\tt MESA} simulation independently predicts a $\approx 1.1\,M_{\odot}$ mass loss after $450$ days, supporting a causal link between the stellar-evolution and envelope-ejecta models (Appendix \ref{app:mesa}).

Upon emerging from the inner boundary, the ejecta expands, radiates, and cools. Because the fitting solutions evolve similarly, we illustrate the low-speed ejecta in Figure \ref{fig:evolution}. Temperature contours at $T=\{1500,3000,6000\}\,$K trace the gas's physical state. The $T=6000\,$K contour proxies the recombination front and photon emission sphere (Appendix \ref{app:radmc3d}). Inside this front, hydrogen bound-free and free-free transitions create a high opacity that lowers the local conventional Eddington luminosity by a factor of $\sim 20$ \citep{chen2026a}. Radiative diffusion within this front (conventionally sub-Eddington; e.g., Figure \ref{fig:evolution}c, white contour) still provides sufficient radiation pressure to accelerate sub-escape ejecta to a super-escape state \citep{chen2024,chen2026a}. Outside the photon emission sphere, the opacity between $6000\,$K and $3000\,$K is dominated by H$^{-}$ absorption—which reprocesses the spectral energy distribution (SED)—and the scattering of short-wavelength photons by newly recombined neutral hydrogen. This shifts the emergent spectrum toward longer wavelengths, causing the blackbody radius to deviate significantly from the conventional $\tau=2/3$ photosphere (Figure \ref{fig:evolution}b). We detail this radiative transfer process in Section \ref{sec:sed}. Below $3000\,$K, molecular line absorption dominates the opacity, assuming no dust. The distant $1500\,$K contour suggests that nearby dust ($\sim 20\,$au) may have sublimated during the outburst.

Measuring the escaping mass based strictly on kinetic energy (Figure \ref{fig:evolution}d) reveals that only $\approx 0.57\,M_{\odot}$ ultimately escapes the binary. The remaining material loses momentum to gas and radiation pressure gradients and falls back toward the envelope. The ejecta velocity distribution evolves over time (Figure \ref{fig:evolution}e). The low-speed ejecta deposits more mass back onto the impenetrable inner boundary, where it piles up, whereas the high-speed ejecta takes longer to fall back.

\section{Ejecta SED modeled with {\tt RADMC-3D}}\label{sec:sed}

\begin{figure*}[!htb]
    \centering
    \includegraphics[width=\textwidth]{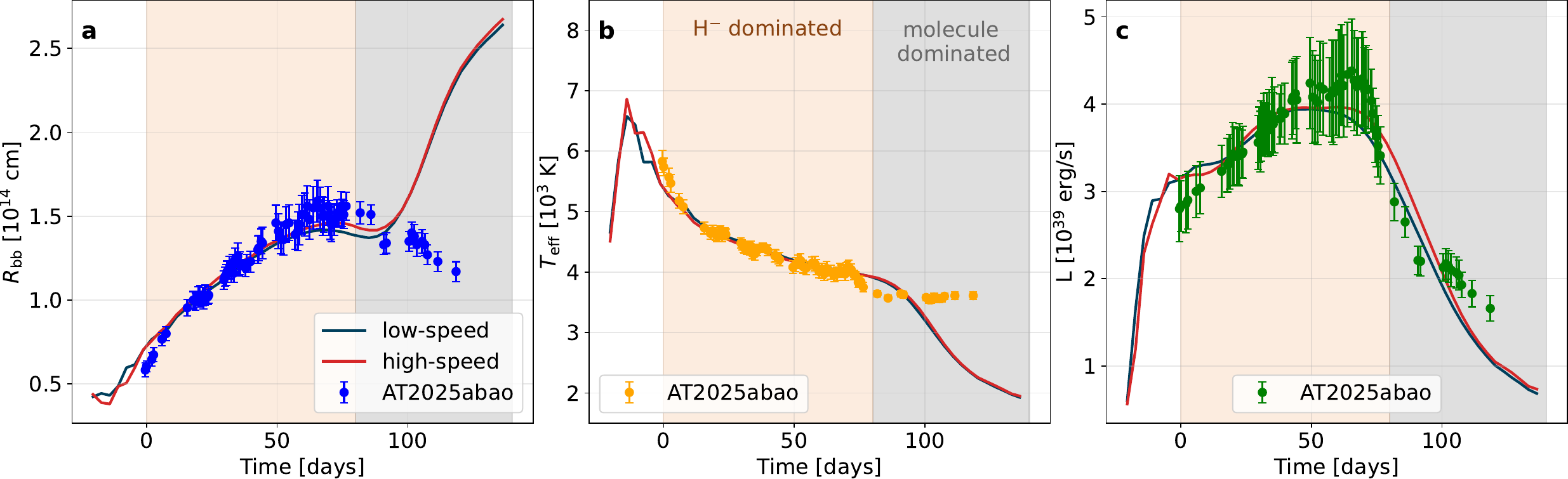}
    \caption{Simulations of the low-speed (blue lines) and high-speed (red lines) ejecta compared with the observations. The simulation times are shifted to match the observed bolometric light curve. Panels a, b, and c: best-fit evolution of $R_{\rm bb}$, $T_{\rm eff}$, and $L$ from the multi-wavelength photometry and from radiative transfer post-processing of the RHD results with {\tt RADMC-3D} (see Appendix \ref{app:radmc3d} for details).}
    \label{fig:lcfitting}
\end{figure*}

One of the advantages of time-domain multi-wavelength photometry is its ability to yield a time-dependent SED, which can be used to deduce the evolution of the effective temperature $T_{\rm eff}$ and blackbody radius $R_{\rm bb}$. The SED depends sensitively on multiple physical conditions, such as the SED of the irradiating source and the scattering and absorption opacities, and therefore provides a stringent test of the ejecta model. To examine the consistency and plausibility of the radiation-pressure-driven ejecta, we post-process the aforementioned two representative solutions with {\tt RADMC-3D} \citep{radmc3d} and compare their $T_{\rm eff}$ and $R_{\rm bb}$ with the observational data of AT\,2025abao. In LRN simulations, the photon-emitting sphere ($r_{\rm em}$) typically has the hydrogen recombination temperature $\approx6000\,$K. In the RHD simulations, radiation is almost in the free-streaming limit outside $r_{\rm em}$ (Figure \ref{fig:evolution}c); we therefore approximate $r_{\rm em}$ by finding a blackbody spherical shell near the recombination front that satisfies $4\pi\sigma r_{\rm em}^{2}T^{4}(r_{\rm em})=4\pi r_{\rm out}^{2}F_{\rm rad}(r_{\rm out})$. The irradiation source is a blackbody of temperature $T_{\rm em}$ and radius $r_{\rm em}$ (Appendix \ref{app:radmc3d}).

At $T<6000\,$K, wavelength-dependent opacities split into two regimes: broad-band H$^{-}$ opacity ($\lambda < 1.6\,\mu$m) and molecular line absorption. We adopt a simplified treatment, calculating the scattering opacity from the RHD $n_{\rm HI}$ and $n_{\rm H_2}$ profiles (Appendix \ref{app:radmc3d}), and using the RHD Rosseland mean opacity as a grey absorption opacity. This grey approximation holds well for smoothly varying opacities like H$^{-}$. However, it breaks down in the molecular regime, where opacity peaks strongly across multiple discrete wavelengths.

Using this irradiating source, alongside the opacities and RHD density profiles, we perform Monte Carlo radiative transfer calculations with {\tt RADMC-3D} to generate synthetic SEDs (Appendix \ref{app:radmc3d}), from which we extract the best-fitting $R_{\rm bb}$ and $T_{\rm eff}$. Figure \ref{fig:lcfitting}a--c presents the resulting evolution of $R_{\rm bb}$, $T_{\rm eff}$, and luminosity for the two RHD solutions. Despite our simplified opacity treatment, the synthetic $R_{\rm bb}$ and $T_{\rm eff}$ agree reasonably well with observations in the H$^{-}$-dominated regime, validating our ejecta mass-loss history estimates. Conversely, fits in the molecular-dominated regime are poor, an expected limitation of applying a grey absorption approximation to discrete molecular lines.

\section{Discussions and Conclusions}

Archival optical and infrared data spanning back to 2003 \citep{law2009,ofek2012,magnier2013,laher2014,cao2016,chambers2016,khan2017,flewelling2020,neugent2020} provide a robust foundation for modeling the recent LRN AT\,2025abao in M31. These data constrain the AGB progenitor to $M_{1}\approx7\pm2\,M_{\odot}$, $\log_{10}(L_{\rm AGB}/L_{\odot})\approx4.2\pm0.1$, and $R\approx350\pm50\,R_{\odot}$ \citep{karambelkar2025}. Furthermore, multi-wavelength photometry of AT\,2025abao \citep{reguitti2026b} and its precursor \citep{karambelkar2025} from 2018 to 2026 offers unique insights into the system's evolutionary sequence: from orbital decay in a strong wind, to CEE, to the LRN outburst. Although direct observational confirmation of a companion is lacking, the strong agreement between theoretical models and long-term data indicates a convergence of observation and CEE theory.

The slow inspiral, governed by the drag-in-the-wind model, naturally explains the gradual brightening observed since 2018 (Figure \ref{fig:precursor}), pointing to a main-sequence companion of $M_{2}\approx1.4^{+0.2}_{-0.3}\,M_{\odot}$ (Figure \ref{fig:precursor}). The subsequent LRN outburst---likely driven by a strong shock launched during the companion's rapid CEE inspiral \citep{gagnier2026}---manifests most prominently in the outer envelope layers. By optimizing a parameterized RHD model (Appendix \ref{app:lcfit}), we find that a continuous spectrum of ejecta velocity profiles—bounded by the low- and high-speed representative solutions—can successfully fit the light curve. This degeneracy arises from the acceleration driven by radiation pressure. In the low- and high-speed ejecta, both solutions predict a similar total outflow mass ($\approx0.92\,M_{\odot}$) and energy ($\approx1.25\times10^{47}\,$erg) (Figure \ref{fig:boundcond}). Gas and radiation pressure gradients strongly accelerate the sub-escape ejecta until approximately $0.57\,M_{\odot}$ of material reaches escape velocity, while the remainder falls back (Figure \ref{fig:evolution}).

During expansion and cooling, the ejecta's deeply embedded recombination front serves as the primary photon emission source for the LRN (Figure \ref{fig:evolution}). Outward-propagating photons are scattered by H and H$_2$ and absorbed by H$^{-}$ and molecules. The resulting opacity reprocesses the radiation, shifting the emergent SED toward longer wavelengths. Both the multi-wavelength photometric fits and the {\tt RADMC-3D} radiative transfer simulations of our RHD solutions consistently capture the observed reddening (Figure \ref{fig:lcfitting}).

Following the CEE phase, the AT\,2025abao system likely consists of the AGB core and the companion in an eccentric orbit ($a\approx10.3\,R_{\odot}$ and $e\approx0.476$ at 400 days). The binary is embedded within a low-density inner radiative layer, yielding a weak gravitational drag. The resulting eccentricity is higher than in some 3D simulations \citep{vetter2025}, reflecting a generally less dissipative environment for angular momentum in the {\tt MESA} model. This discrepancy may arise because our 1D model lacks several dissipative mechanisms captured by 3D models—such as angular momentum transport via dynamo-amplified MRI \citep{gagnier2024,vetter2025} and strong dynamical tides at periastron \citep{lai1997}—which would otherwise circularize the orbit. However, we caution that the distinct initial conditions and energy transport mechanisms between 1D and 3D models can also contribute to the differences in final eccentricities.

If these combined angular momentum transfer mechanisms sufficiently tighten the binary, the low-density helium envelope will likely be levitated. Concurrently, the hydrogen envelope expands to $\gtrsim6\,$au (Figure \ref{fig:cee1d}). If AT\,2025abao is a hierarchical system, outer tertiary objects (within $\sim10\,$au) could interact with this expanded envelope, further assisting in unbinding it \citep{hamers2022,shariat2025,xu2026}. If the envelope is not ejected before the companion reaches its tidal disruption radius ($\approx3.38\,R_{\odot}$), the companion will be tidally disrupted. Otherwise, the system will survive as a close main-sequence star and C/O white dwarf (WD) binary. Billions of years later, as the main-sequence star evolves, the WD will strip its envelope, leaving a helium core that will become a hot subdwarf \citep{han2002,lin2023}.

AT\,2025abao may represent the first observed CEE event to form a close compact binary. Interpreting this system relies critically on the long-term optical and infrared baseline surveys that characterized its progenitor. Furthermore, rapid transient identification and follow-up multi-wavelength photometry are essential for constraining ejecta masses. In the era of the Vera C. Rubin Observatory's Legacy Survey of Space and Time (LSST) \citep{lsst,lsst2}, an expected yield of $\sim100$ LRNe per year \citep{howitt2020} will likely include many evolved systems on the path to becoming compact binaries. Combining these transient discoveries with deep multi-wavelength stellar surveys \citep{dalcanton2009,boyer2015,mcquinn2017} and 3D RHD simulations \citep{goldberg2022,lau2025a,ma2025} will rapidly advance our understanding of CEE, LRNe, and compact binary formation \citep{han2020,chenxuefei2024}.

\appendix
\section{Photometry} \label{subsec:photometry}

The photometric data of AT\,2025abao presented in this paper were obtained from \citet{reguitti2026b} alongside their subsequent late-time observations. In addition, we included optical light curves in the Sloan $ugri$ filters from the Wide Field Survey Telescope \cite[WFST;][]{wang2023,cai2025,lin2025}. WFST is a 2.5-m-aperture wide-field telescope located at the Lenghu Observatory \citep{deng2021} and equipped with a 0.73-gigapixel mosaic CCD camera. The WFST data were reduced using a dedicated data-reduction pipeline based on a modified version of the LSST Science Pipelines \citep{juric2017,cai2025}. The reduction included the removal of instrumental signatures using bias, flat-field, fringe, and crosstalk calibration frames, followed by the identification of pixels affected by blending, saturation, cosmic rays, and other artifacts. The sky background was modeled using a two-dimensional sixth-order Chebyshev polynomial and subtracted from the science images. Sources were then detected on the background-subtracted images, and their fluxes were measured using point-spread-function photometry. The astrometric solution was calibrated against Gaia DR3 \cite[][]{gaia2016,gaia2022}, while the photometric parameters were calibrated using the Sloan Digital Sky Survey for the $u$ band \citep{fukugita1996,york2000} and Pan-STARRS DR2 for the $gri$ bands \citep{chambers2016,flewelling2020,magnier2020}. 
%The resulting photometric measurements were converted to AB magnitudes, and the exposure midpoints were converted into  barycentric modified Julian dates with barycentric dynamical time ($\rm BMJD_{TDB}$). 
We note that measurements affected by blending, saturation, cosmic-ray contamination, or proximity to the image boundary were excluded, and the calibrated measurements were cross-matched within a radius of $1^{\prime\prime}$ \citep{lin2025} to obtain the WFST photometric series for AT\,2025abao. 
Finally, the magnitudes of AT\,2025abao are available in the public repository\footnote{The zenodo repository at \dataset[doi: 10.5281/zenodo.23175368]{https://doi.org/10.5281/zenodo.23175368} contains the WFST photometry data used in this work.}.

Adopting the same distance ($\mu = 24.47\pm0.06$~mag) and extinction ($A_{\rm V} =0.17$~mag) as \citet{reguitti2026b}, we derive an updated bolometric luminosity (L$_{\rm BB}$) for AT\,2025abao through a single-component blackbody (BB) fit. Following the methodology of \citet{reguitti2026b}, this fit yields the blackbody temperature (T$_{\rm BB}$) and the radius (R$_{\rm BB}$). The updated evolution of L$_{\rm BB}$,  T$_{\rm BB}$, and R$_{\rm BB}$ is shown in Figure~\ref{fig:lcfitting}.

\section{Drag-in-the-wind model}\label{app:wind}

We consider the binary in a lab frame with the origin located at the binary's center of mass. The AGB star launches a strong wind under the influence of the companion while rotating and orbiting. When the companion star travels through the wind, it imposes a gravitational drag force \citep{ostriker1999} on the supersonic wind, which is not considered part of the primary star. The equations of motion of the binary are:
\begin{eqnarray}
    \mathbf{\Ddot{x}}_{\rm AGB}=\mathbf{\Ddot{x}}_{1}&=&\frac{F_{g}}{M_{1}}\mathbf{\hat{a}}_{\rm{orb}},    \\
    \mathbf{\Ddot{x}}_{2}&=&-\frac{F_{g}}{M_{2}}\mathbf{\hat{a}}_{\rm{orb}}-\frac{F_{\rm d}}{M_2}\mathbf{\hat{v}}_{\rm{rel}},
\end{eqnarray}
where $\mathbf{\hat{a}}_{\rm{orb}}=(\mathbf{x}_{2}-\mathbf{x}_{1})/\|\mathbf{x}_{2}-\mathbf{x}_{1}\|$, $F_{g}=GM_{1}M_{2}/a_{\rm{orb}}^{2}$, $\mathbf{\hat{v}}_{\rm{rel}}$ is the unit relative velocity between the companion and the wind, and $F_{\rm d}$ is the gravitational drag force \citep{ostriker1999,kim2007,kim2010}; here we adopt the same formulas as in \citet{bronner2024}, with an impact radius parameter $C_{d}$. The gas in the wind gains angular momentum from the companion due to the drag force. In the wind interaction regime, all the angular momentum lost through the drag force is carried away by the wind mass loss. The total angular momentum loss rate is given by:
\begin{eqnarray}
    \dot{\mathbf{J}}_{\rm total}&=&\underbrace{\bigg(\Omega_{\rm env,i}R_{\rm rot}^{2}+\frac{M_{2}^{2}\sqrt{Ga_{\rm orb}}}{(M_{1}+M_{2})^{3/2}}\bigg)\dot{M}_{1}\mathbf{\hat{z}}}_{\text{AM loss from the AGB}} \nonumber\\
    &&+\underbrace{F_{\rm d}\mathbf{x}_2\times\mathbf{\hat{v}}_{\rm{rel}}}_{\text{companion}},
\end{eqnarray}
where we assume the wind from the primary star initially carries the rotational and orbital specific angular momentum from the AGB star, with $\Omega_{\rm env,i}\approx8.82\times10^{-8}\,\rm{rad\,s^{-1}}$ and $R_{\rm rot}\approx380\,R_{\odot}$ defined in Appendix \ref{app:mesainit}. Additionally, we assume the angular momentum loss rate from the drag force is limited by the scenario in which all the mass leaving the binary system carries the specific angular momentum of the L2 point; the drag force in the wind is therefore capped at:
\begin{eqnarray}
    F_{\rm d}&\le&\bigg(\sqrt{G(M_{1}+M_{2})r_{L2}}-\Omega_{\rm env,i}R_{\rm rot}^{2} \nonumber\\
    &&-\frac{M_{2}^{2}\sqrt{Ga_{\rm orb}}}{(M_{1}+M_{2})^{3/2}}\bigg)\frac{\dot{M}_{1}}{a_{\rm orb}},   \label{eqn:fdlimit}\\
    r_{L2}&\approx&(1+0.75q^{1/3}-0.47q^{2/3}-0.09q)a_{\rm orb},
\end{eqnarray}
where $q=M_{2}/M_{1}$ is the mass ratio. This upper limit (Equation \ref{eqn:fdlimit}) becomes the effective drag force formula when $F_{\rm d}\propto C_{d}(M_{2}/v_{\rm{rel}})^{2}$ given by the formulas in \citet{bronner2024} is larger.

The physical condition of the wind around the AGB star in the presence of a close companion can be very complicated. Studies show that the extended AGB atmosphere can pulsate. However, the terminal velocity of the wind usually falls in the range of $[20,30]$\,km\,s$^{-1}$ in high-mass AGB stars \citep{hofner2018,decin2021}. From a time-averaged perspective, we approximate the AGB wind with a steady-state wind at $v_{\rm wind}=25$\,km\,s$^{-1}$, and the density profile and the relative velocity are approximated by,
\begin{eqnarray}
    \mathbf{v}_{\rm{rel}}&=&\mathbf{\dot{x}}_{2}-\mathbf{\dot{x}}_{1}-\frac{\Omega_{\rm{env,i}}R_{\rm{rot}}^2}{a_{\rm{orb}}}\mathbf{\hat{\phi}}-v_{\rm wind}\mathbf{\hat{r}},    \\
    \rho_{\rm wind}&=&\dot{M}_{1}/(4\pi a_{\rm orb}^{2}v_{\rm wind}),
\end{eqnarray}
where $\dot{M}_{1}=10^{-3}\exp(t/(3\rm{yr}))\,M_{\odot}\,\rm{yr}^{-1}$ is the prescribed mass-loss rate, mimicking a runaway mass-loss history as the binary gets closer in \citep{ge2020,temmink2023}. The initial binary separation is set at $a_{\rm orb}=480\,R_{\odot}$. At this separation, increasing/decreasing $a_{\rm orb}$ will mainly change the orbital evolution (expand/shrink the yellow region in Figure \ref{fig:precursor}) and energy budget of $\dot{E}_{\rm heat}$ (shift downward/upward the pink region). The overall blue region remain focused around $M_{2}\approx1.4^{+0.2}_{-0.3}\,M_{\odot}$.

Finally, a companion star in the AGB wind can strongly deflect the wind toward the companion, enhancing $\rho_{\rm wind}$ \citep{mohamed2007,chen2017,chen2020,tejeda2025,huang2026}. We adopt $C_{d}\in\{1.5,2,2.5\}$ in this model to account for the impact of the focused wind. Because $F_{\rm d}\propto C_{d}$, varying $C_d$ can change the angular momentum loss limited parameter space, denoted by $\times$ in Figure \ref{fig:precursor}. We assume the work done by the drag force is converted into radiated luminosity,
\begin{equation}
    \dot{E}_{\rm heat}=-F_{\rm d}\,\hat{\mathbf{v}}_{\rm rel}\!\cdot\!\dot{\mathbf{x}}_{2},
\end{equation}
and the total luminosity is $L_{\rm total}=L_{\rm AGB}+\dot{E}_{\rm heat}$.

\section{Inspiral model in the envelope}\label{app:mesa}

We model the common envelope inspiral in a 1D {\tt MESA} (r10398) run by coupling the stellar structure to an externally integrated binary orbit\footnote{The zenodo repository at \dataset[doi: 10.5281/zenodo.23191838]{https://doi.org/10.5281/zenodo.23191838} contains the information about the 1D CEE MESA model we use in this work.}, similar to \citet{fragos2019,bronner2024}. The envelope is treated as concentric shells that rotate about the $z$ axis and co-move with the primary core, so the core feels no friction from its own envelope. The companion (mass $M_2 = 1.4\,M_\odot$) orbits at separation $\aorb$; the primary is decomposed into a core $M_c$ and envelope $M_{\rm env}$, with $M_{\rm 1,orb}\equiv M(\aorb)$ the primary mass enclosed within the orbit.

\subsection{Initial condition setup with {\tt MESA}}\label{app:mesainit}

The adopted primary star has a zero-age main sequence mass of $M=7\,\rm{M}_{\odot}$ and solar metallicity \citep{asplund2021}. We evolve it to the early AGB phase, reaching $M_{1}=6.9\,\rm{M}_{\odot}$, $R\approx343\,\rm{R}_{\odot}$, and $\log_{10}(L_{\rm AGB}/\rm{L}_{\odot})\approx4.23$. Our main inlist settings follow \citet{choi2016} (see their Section 3 for details).

Assuming the stellar envelope has been spun up by the companion, we construct a rotating envelope model with a rotational frequency below corotation. For reference, the orbital frequency of the binary system is:
\begin{equation}
    \Omega=1.066\times10^{-7}\left(\frac{a_{\rm RLOF}}{a}\right)^{1.5}\,\rm{rad\,s^{-1}},
\end{equation}
where $a_{\rm RLOF}\approx660\,\rm{R}_{\odot}$ is the binary separation at the onset of Roche-lobe overflow, under the assumption of no rotational inflation of the envelope.

The relaxation process increases the stellar radius to $R_{\rm rot}\approx380\,\rm{R}_{\odot}$ with $\Omega_{\rm env,i}\approx8.82\times10^{-8}\,\rm{rad\,s^{-1}}$. Because the exact envelope rotation frequency is difficult to constrain, we adopt this moderate value to illustrate the impact of rotation. Detailed modeling is warranted, as rotation dictates the initial state of the CEE. We use this relaxed, rotating model to initialize our subsequent CEE calculations, setting the initial orbital separation to $a_{\rm orb}=380\,\rm{R}_{\odot}$.

\subsection{Conserved quantities}
Working in the lab frame centered on the center of mass,
\begin{eqnarray}
    \mathbf{p}_{\rm total}&=&M_{1}\dot{\mathbf{x}}_{1}+M_{2}\dot{\mathbf{x}}_{2},\\
    \mathbf{J}_{\rm total}&=&J_{\rm env}\hat{\mathbf{z}}
        +M_{2}\mathbf{x}_{2}\times\dot{\mathbf{x}}_{2}+M_{1}\mathbf{x}_{1}\times\dot{\mathbf{x}}_{1},\\
    E_{\rm total}&=&E_{\rm int}+E_{1}+E_{2}+E_{\rm rot}+E_{\phi},
\end{eqnarray}
with $E_{i}=\tfrac12 M_{i}\dot{x}_{i}^{2}$ the bulk kinetic energy of each body,
$E_{\rm rot}$ the envelope rotational energy, and the gravitational energy
\begin{equation}\label{eqn:gravitation}
    E_{\phi}=\underbrace{-\frac{GM_{\rm 1,orb}M_{2}}{\aorb}}_{\rm interior}
    -\underbrace{\int_{\aorb}^{\infty}4\pi\rho r\,GM_{2}\,dr}_{\rm exterior}
    -\underbrace{\int_{M_c}^{M_{1}}\frac{GM\,dM}{r}}_{\rm self}.
\end{equation}
The interior term uses only the enclosed mass $M_{\rm 1,orb}$, since the
envelope exterior to the orbit exerts no net force on the binary; conversely,
that exterior envelope \emph{does} feel the companion, which we impose on the
stellar model.

\subsection{Orbital evolution}
The two bodies obey
\begin{eqnarray}
    \ddot{\mathbf{x}}_{1}&=&\frac{F_{g}}{M_{1}}\hat{\mathbf{a}}_{\rm orb}
        +\frac{F_{\rm d}}{M_{1}}\hat{\mathbf{v}}_{\rm rel},\label{eqn:kin1}\\
    \ddot{\mathbf{x}}_{2}&=&-\frac{F_{g}}{M_{2}}\hat{\mathbf{a}}_{\rm orb}
        -\frac{F_{\rm d}}{M_{2}}\hat{\mathbf{v}}_{\rm rel},\label{eqn:kin2}\\
    \dot{J}_{\rm env}\hat{\bf{z}}&=&F_{\rm d}\,(\mathbf{x}_{2}-\mathbf{x}_{1})\times\hat{\mathbf{v}}_{\rm rel},
        \label{eqn:angular}
\end{eqnarray}
where $F_{g}=GM_{\rm 1,orb}M_{2}/\aorb^{2}$ and the gas-relative velocity is
\begin{eqnarray}
    &&\mathbf{v}_{\rm rel}=\dot{\mathbf{x}}_{2}-\dot{\mathbf{x}}_{1}
        -\Omega\,\hat{\mathbf{z}}\times\mathbf{a}_{\rm orb}
        -v_{r}\,\hat{\mathbf{a}}_{\rm orb},\\
    &&\mathbf{a}_{\rm orb}=\mathbf{x}_{2}-\mathbf{x}_{1},
\end{eqnarray}
with $\Omega$ and $v_{r}$ the envelope angular frequency and radial
expansion speed at $r=\aorb$. Equation~\eqref{eqn:angular} follows from
$d\mathbf{J}_{\rm total}/dt$ with $\hat{\mathbf{a}}_{\rm orb}\parallel\mathbf{x}_1
\parallel\mathbf{x}_2$: the drag torque removes orbital angular momentum and
deposits it locally in the envelope,
\begin{eqnarray}
    &&\dot{J}_{\rm env}=\int_{r_{\min}}^{r_{\max}}\frac{8\pi\rho r^{4}\dot\Omega}{3}\,dr,\\
    &&\dot{E}_{\rm rot}=\int_{r_{\min}}^{r_{\max}}\frac{8\pi\rho r^{4}\Omega\dot\Omega}{3}\,dr,
\end{eqnarray}
over the same region $[r_{\min},r_{\max}]$ in which heat is deposited. The drag
force $F_{\rm d}(\mathcal{M})$ is the Mach-number-dependent dynamical friction of
\citet{bronner2024} (their Eq.~5), with subsonic/supersonic Coulomb terms and a
nonlinear density enhancement, scaled by $C_d$.

\subsection{Heating from energy conservation}\label{app:ch}
Energy conservation, $\dot{E}_{\rm total}=0$, combined with
Equations~\eqref{eqn:kin1}--\eqref{eqn:angular} gives the net drag heating that must
be added to the envelope,
\begin{equation}\label{eqn:heat}
    \dot{E}_{\rm heat}=\dot{E}_{\rm k}-\dot{E}_{\rm rot}
        -\dot{E}_{\phi,\rm int}-\dot{E}_{\phi,\rm ext},
\end{equation}
where $\dot{E}_{\rm k}=F_{\rm d}\hat{\mathbf{v}}_{\rm rel}\!\cdot\!(\dot{\mathbf{x}}_{2}-\dot{\mathbf{x}}_{1})
+F_{g}\hat{\mathbf{a}}_{\rm orb}\!\cdot\!(\dot{\mathbf{x}}_{2}-\dot{\mathbf{x}}_{1})$
is the work done by drag and gravity on the binary. Using
$dM_{\rm 1,orb}/dt=4\pi\rho(\aorb)\aorb^{2}\dot{a}_{\rm orb}$, the interior and
exterior potential terms are
\begin{eqnarray}
    \dot{E}_{\phi,\rm int}&=&-\frac{GM_{2}}{\aorb}\dv{M_{\rm 1,orb}}{t}
        +F_{g}\hat{\mathbf{a}}_{\rm orb}\!\cdot\!(\dot{\mathbf{x}}_{2}-\dot{\mathbf{x}}_{1}),\\
    \dot{E}_{\phi,\rm ext}&=&+\frac{GM_{2}}{\aorb}\dv{M_{\rm 1,orb}}{t},
\end{eqnarray}
so the interior and exterior gravitational potential and $F_g$ terms cancel and Equation~\eqref{eqn:heat}
reduces to
\begin{equation}\label{eqn:heat_final}
    \dot{E}_{\rm heat}
    =F_{\rm d}\,\hat{\mathbf{v}}_{\rm rel}\!\cdot\!(\dot{\mathbf{x}}_{2}-\dot{\mathbf{x}}_{1})
    -\dot{E}_{\rm rot}
\end{equation}
i.e.\ the drag-force power minus the fraction stored as envelope rotation. The remainder is deposited as heat and redistributed by {\tt MESA} into internal energy and $p\,dV$ work (envelope expansion) within a finite region around the companion. Following \citet{bronner2024}, the radial extent of the heating region is controlled by the dimensionless parameter $C_{\rm h}$, which scales the heating region relative to the local Bondi--Hoyle--Lyttleton accretion radius. In our calculations, we adopt $C_{\rm h}=1$. This energy deposition approximation is likely more diffusive than physical shocks. Future 2D/3D RHD simulations are needed to better constrain $C_{\rm h}$.

\subsection{Energetics and mass loss}

We analyze the energetics of the plunge-in phase by comparing the energies of the initial CEE state at $t_{\rm i}$ with those at the onset of the self-regulated phase at $t_{\rm f}\approx200$ days. We decompose the energy into six components to better understand the energy conversion during the plunge-in phase.
\begin{itemize}[leftmargin=15pt]
    \item $\Phi_{\rm orbit}=-GM_{\rm 1,orb}M_{2}/\aorb$ is the gravitational potential energy between the two stars.
    \item $E_{\rm int}$ is the total internal energy of the envelope.
    \item $\Phi_{\rm env}$ is the envelope gravitational potential energy, i.e., the sum of the ``exterior" and ``self" parts in Equation \ref{eqn:gravitation}.
    \item $E_{k}$ is the kinetic energy of the two stars, including the envelope.
    \item $E_{\rm rot}$ is the envelope rotational energy.
    \item $\Delta E_{\rm L}=\int_{t_{\rm i}}^{t_{\rm f}}(L_{\rm photo}-L_{\rm nuc})dt$ is the excess energy radiated away before the onset of the self-regulated phase.
\end{itemize}

\begin{figure*}[!htb]
    \centering
    \includegraphics[width=\textwidth]{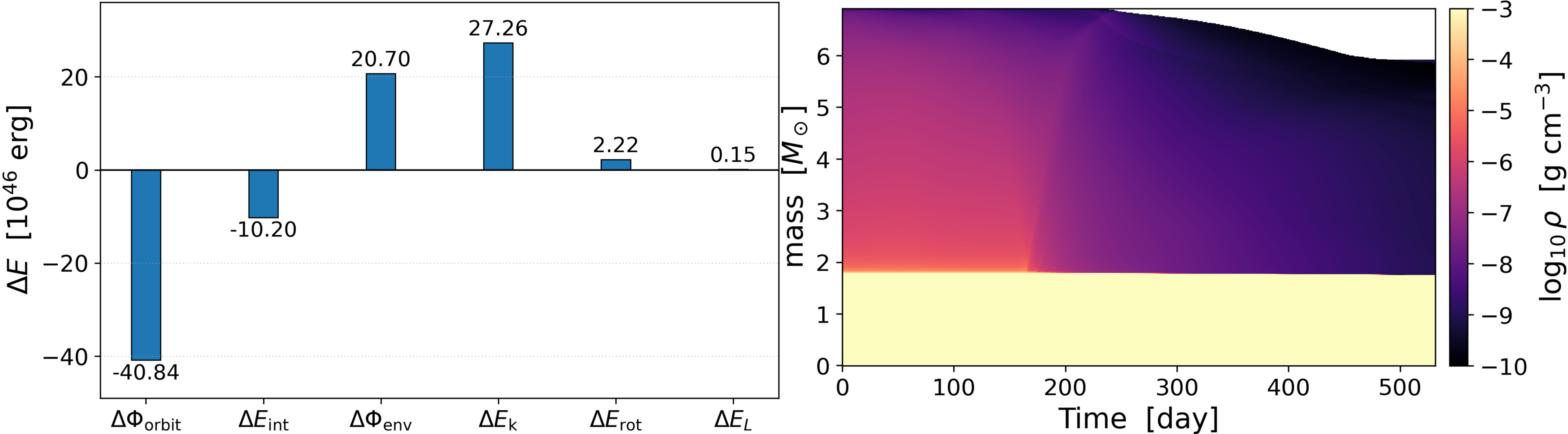}
    \caption{Left: the energy differences between the average from day 180 to 200 and the initial state for the six decomposed energy components. Right: Kippenhahn diagram of the density showing the mass loss and significant envelope expansion.}
    \label{fig:energetics}
\end{figure*}

Because the companion is on an eccentric orbit, we average the energies between 180 and 200 days and subtract the energies of the initial state from the averaged values, as shown in Figure \ref{fig:energetics}. The orbital gravitational potential energy and internal energy are converted into the envelope gravitational potential energy and kinetic energy. The envelope is also spun up due to angular momentum transfer and gains some rotational energy. The levitated envelope consumes a large fraction of the converted energy \citep{temmink2023,temmink2025}. The excess luminosity during this plunge-in phase is only $\approx1.5\times10^{45}\,$erg, equivalent to an average luminosity of $22000\,L_{\odot}$ over 200 days. This is even smaller than the largest $\dot{E}_{\rm heat}$ during the drag-in-the-wind phase. Given that the envelope is usually reddened or obscured by dust during this pre-outburst phase \citep{macleod2022}, it could be fainter in the optical, making it ``dimmer" than the precursor \citep{tylenda2011}.

As the self-regulated evolution begins, the envelope may pulsate at super-escape speeds. We remove the super-escape envelope with a wind in {\tt MESA}. The Kippenhahn diagram in Figure \ref{fig:energetics} shows that $\approx1.1\,M_{\odot}$ of the envelope is lost before the whole envelope settles into a sub-escape state.

\section{Light curve fitting with {\tt Guangqi}}\label{app:lcfit}

To find the outflow parameters that reproduce the observed light curve of AT\,2025abao, we perform an automated inverse-modeling search over the ten scalar boundary condition parameters governing the (radial) ejecta velocity $v(t)$, mass injection rate $\dot{M}(t)$, and temperature $T(t)$ histories, with the central mass and inner radius held fixed. The specific parameterized functional forms of $v$, $\dot{M}$, and $T$ are complex (e.g., Appendix C of \citet{chen2024}), but their exact mathematical expressions are not critical. The essential requirement is that they produce a bump before asymptotically decaying to negligible values. Therefore, any sufficiently flexible function exhibiting this behavior can yield valid boundary conditions that qualify as ``solutions''.

We evaluate each candidate set of $v(t)$, $\dot{M}(t)$, and $T(t)$ by running a full RHD simulation with the {\tt Guangqi} code \citep{chen2026b}. The goodness of fit is quantified by the weighted mean fractional deviation of the modeled bolometric luminosity from the observations:
\begin{equation}
\mathcal{S} = \frac{\sum_i w_i\bigl|L_{\rm sim}(p_i) - L_{\rm obs}(p_i)\bigr|
/ L_{\rm obs}(p_i)}{\sum_i w_i},
\end{equation}
evaluated at the observed phases $p_i$ after a one-dimensional search for the optimal temporal alignment of the model. The weight was unity for $p_i \le 70$~d and decayed exponentially, $w = 0.5 + 0.5\,e^{-(p_i-70)/\tau}$ with $\tau = 10$~d thereafter, so as to emphasize the rise and peak over the photometrically noisier tail. Although an individual simulation is inexpensive ($\approx$6 min), an exhaustive search of the ten-dimensional space is infeasible, and the objective is a gradient-free, stochastic black box; we therefore adopt sample-efficient Bayesian optimization. An initial Latin-hypercube design \citep{mckay1979} provides an unbiased global sampling of the response surface, after which a GP surrogate \citep{rasmussen2006} with a Mat\'ern-5/2 kernel is iteratively refined and queried using an EI acquisition function \citep{jones1998}. Multiple candidates are proposed per iteration through the constant-liar heuristic \citep{ginsbourger2010} to exploit parallel execution across cores. The Gaussian-process regression and acquisition are implemented with {\tt scikit-learn} \citep{pedregosa2011} and the Latin-hypercube sampler with {\tt scipy} \citep{virtanen2020}.

We carried out numerous independent searches and present two representative solutions in Figure \ref{fig:boundcond}. The low-speed ejecta has higher $E_{r}/\eg$ and $p_{r}/p_{g}$ than the high-speed ejecta because it requires more radiation pressure to be accelerated to sub-escape speeds. Overall, different solutions vary in the shape of $\rho$, $v_{\rm ej}$, and $T$, but have similar total amounts of mass and energy injection ($\Delta M\approx0.92\,M_{\odot},\ E_{\rm tot}\approx1.25\times10^{47}\,\rm{erg}$), implying robust estimates of the mass loss and energetics.

\section{Radiative transfer with {\tt RADMC-3D}}\label{app:radmc3d}

To convert the grey radiation-hydrodynamic (RHD) output into a synthetic SED and to extract the blackbody temperature and radius that observers report, we post-process each simulation snapshot with the publicly available Monte-Carlo radiative-transfer code {\tt RADMC-3D}. The calculation is performed on a 1D spherical grid that inherits the radial cell structure of the {\tt Guangqi} RHD run, one {\tt RADMC-3D} model per output time.

\subsection{Central source}
Each snapshot provides, at the outer boundary $r_{\rm out}$, the bolometric
luminosity
\begin{equation}
  L \;=\; 4\pi\, r_{\rm out}^{2}\, F_{\rm rad}(r_{\rm out}),
\end{equation}
where $F_{\rm rad}$ is the \emph{net} (outward minus inward) radiation flux from the grey RHD calculation, so that $L$ already accounts for the cancellation between the outgoing and returning photons and represents the true energy loss rate of the envelope. {\tt RADMC-3D}, however, requires a source that emits purely outward. We therefore define an inner emitting shell as the radius $r_{\rm em}$ at which the local blackbody luminosity matches $L$,
\begin{equation}
  4\pi\, r_{\rm em}^{2}\, \sigma\, T_{\rm em}^{4} \;=\; L ,
  \label{eq:innershell}
\end{equation}
with $T_{\rm em}=T(r_{\rm em})$ the local radiation temperature, and assume that all of its flux propagates outward. This shell is used as the illuminating source: a blackbody of temperature $T_{\rm em}$ is placed just at $r_{\rm em}$, and its emergent intensity is propagated through the overlying envelope so that scattering and reprocessing shape the final SED.

\subsection{Opacities}
During the outburst, the computational domain contains essentially no dust, so
the continuum opacity is set by the gas. For \emph{absorption}, we adopt the
grey Rosseland mean opacity $\kappa_{\rm R}$ used in the RHD simulation, which
combines the atomic gas, H$^{-}$, and molecular contributions. Because the
blackbody photosphere generally lies at or beyond the recombination front, the
gas there is atomic or molecular, and H$^{-}$ provides an efficient
broad-band opacity. We reproduce the grey absorption optical depth exactly by
folding $\kappa_{\rm R}$ into an effective absorber density: with a
frequency-independent opacity $\kappa_{\rm abs}=10^{-3}\,\mathrm{cm^2\,g^{-1}}$,
we assign a density $\rho_{\rm abs}=\rho\,\kappa_{\rm R}\times10^{3}$, so that
\begin{equation}
  \tau_{\rm abs}
  = \int \rho_{\rm abs}\,\kappa_{\rm abs}\,\mathrm{d}r
  = \int \rho\,\kappa_{\rm R}\,\mathrm{d}r ,
\end{equation}
i.e.\ the {\tt RADMC-3D} absorption optical depth is identical to the grey RHD one
and remains wavelength-independent. For \emph{scattering}, we include the
classic wavelength-dependent Rayleigh scattering off neutral hydrogen (H\,\textsc{i})
and molecular hydrogen (H$_2$), with number densities $n_{\rm{HI}}$ and $n_{\rm{H}_2}$ taken from the RHD simulations, and consider isotropic scattering.

\subsection{SED and blackbody fit}

\begin{figure*}
    \centering
    \includegraphics[width=\textwidth]{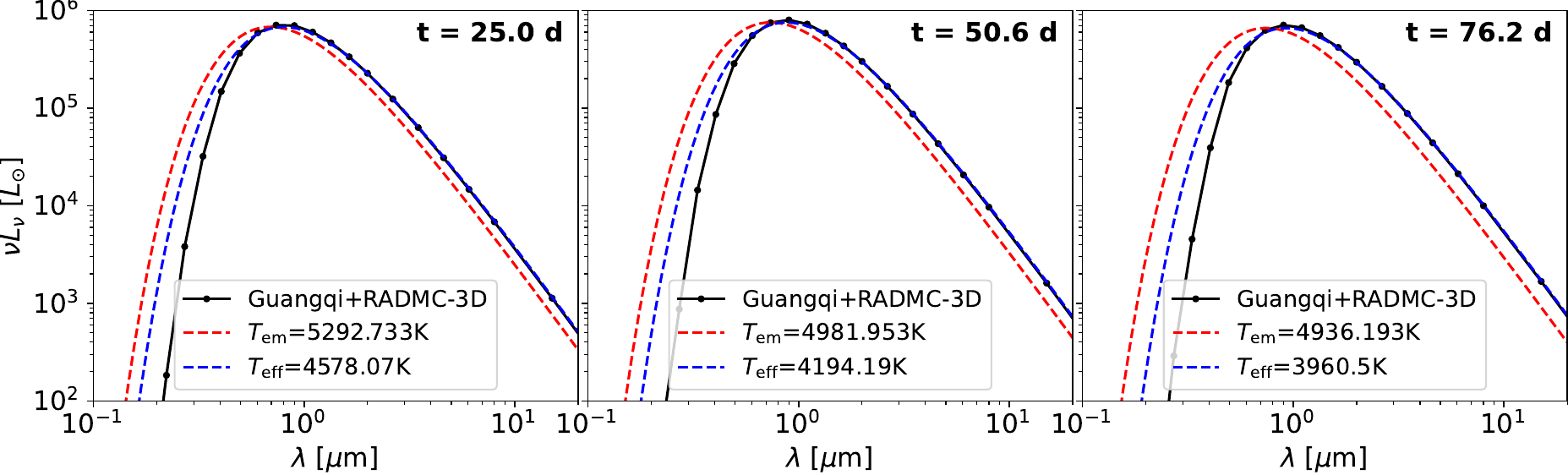}
    \caption{Spectral luminosities of the low-speed ejecta at different times. The red dashed line indicates the effective temperature of the photon emission sphere at $r_{\rm em}$. The black dotted line shows the spectral luminosity obtained from radiative transfer post-processing of the {\tt Guangqi} RHD solution with {\tt RADMC-3D}. The blue dashed line shows the best-fit blackbody.}
    \label{fig:sed}
\end{figure*}

For each snapshot we run a thermal Monte-Carlo calculation followed by an SED
calculation, yielding the emergent monochromatic luminosity $L_{\nu} = 4\pi r_{\rm out}^{2} F_\nu$ at 25 wavelengths ranging from 0.1 to 100 $\mu$m. We present the spectral luminosities of the low-speed ejecta case at different times as an example in Figure \ref{fig:sed}. We determine the effective temperature by fitting a single-temperature blackbody to the SED with the {\tt curve\_fit} function from {\tt scipy} \citep{virtanen2020}. The corresponding blackbody radius is then fixed by requiring the fitted blackbody to radiate the snapshot luminosity, $R_{\rm bb}=\sqrt{L/(4\pi\sigma T_{\rm eff}^{4})}$, which is directly comparable to the $R_{\rm bb}$ obtained from observed SEDs. An example fit is shown in Figure \ref{fig:sed}: the synthetic SED (black) is matched closely by the fitted blackbody (blue), which is slightly cooler than the emitting-shell temperature $T_{\rm em}$ (red) because scattering and line-blanketing redistribute flux from the blue toward longer wavelengths.

\subsection{Validity}
The recovered $T_{\rm eff}$, $R_{\rm bb}$, and $L$ track the observations well
during the plateau phase
(Figure~\ref{fig:lcfitting}, $0\lesssim\rm{Time}\lesssim 80\,$d). Throughout this phase the
photospheric temperature is high enough that H$^{-}$ is the dominant continuum
opacity source; its smooth, broad-band character makes the grey approximation
accurate, so the reprocessed SED remains close to a blackbody and the fit is
robust. At late times, once the photospheric temperature drops below
$\approx3000\,$K (but stays above the dust-condensation temperature), molecular
bands become the dominant opacity. This opacity is line-dominated and strongly
frequency-dependent, so the grey treatment inherited from the RHD run is no
longer a good approximation: the model $T_{\rm eff}$ falls well below the
observed values, which flatten near $\approx3600\,$K, while the inferred
$R_{\rm bb}$ correspondingly diverges upward (Figure~\ref{fig:lcfitting},
$\rm{Time}\gtrsim100\,$d). The blackbody fit should therefore be regarded as reliable
only up to the onset of the molecular-dominated regime.

\section*{Contribution}

Zhuo Chen led the writing of the manuscript, proposed the drag-in-the-wind and CEE inspiral models, and performed the GP and EI optimizations with \texttt{Guangqi} as well as the radiative-transfer simulations with \texttt{RADMC-3D}. Yingzhen Cui implemented the CEE model in \texttt{MESA} and carried out most of the \texttt{MESA} simulations. Tianshu Wu contributed to the light-curve fitting with \texttt{Guangqi}. Andrea Reguitti and Yongzhi Cai performed the photometric analysis; Yongzhi Cai drafted the corresponding sections of the manuscript. Hongwei Ge, Jie Lin, and Lulu Fan contributed to the scientific discussions and provided valuable insights. Minxuan Cai, Lulu Fan, Ji-An Jiang, Tinggui Wang, Zelin Xu, and Jie Lin are members of the WFST team and provided the observational data.

\begin{acknowledgments}
The authors thank the anonymous referee for the highly constructive suggestions. The authors also thank Viraj Karambelkar for sharing the light curve of WNTR23bzdiq/WTP19aalzlk. The Wide Field Survey Telescope (WFST) is a joint facility of the University of Science and Technology of China and the Purple Mountain Observatory.
This work is supported by National Key R\&D Program of China (Grant No.~2025YFF0511003), the National Natural Science Foundation of China under grants No.~12473030, 12103028, 12303054, 12342501, Tsinghua University Dushi Program, the National Key R\&D Program of China (Grant No.~2024YFA1611603), the Yunnan Fundamental Research Projects (Grant Nos. 202401AU070063, 202501AS070078). Y.-Z.C. and A.R. acknowledge financial support from the SOXS project (PI S.~Campana). A.R. also acknowledges financial support from the PRIN-INAF 2022 ``Shedding light on the nature of gap transients: from the observations to the models''.
J. L. is supported by the National Natural Science Foundation of China (Grant No. 12403038), the Fundamental Research Funds for the Central Universities (Grant No. WK2030000089), and the Cyrus Chun Ying Tang Foundation. The Center of High Performance Computing at Tsinghua University provided the computational resources.
\end{acknowledgments}

\facilities{WFST: 2.5m}

\software{{\tt MESA} \citep{mesa2011,mesa2013,mesa2015,mesa2018,mesa2019,mesa2023},
          {\tt Guangqi} \citep{chen2026b},  
          {\tt RADMC-3D} \citep{radmc3d},
          {\tt scipy} \citep{virtanen2020}
          }

\bibliography{references}
\bibliographystyle{aasjournalv7}

%% This command is needed to show the entire author+affiliation list when
%% the collaboration and author truncation commands are used.  It has to
%% go at the end of the manuscript.
%\allauthors

%% Include this line if you are using the \added, \replaced, \deleted
%% commands to see a summary list of all changes at the end of the article.
%\listofchanges
\end{CJK*}
\end{document}